


\input harvmac
\def\Oops{(2.2)}


\overfullrule=0pt


\def\as{\alpha_s}
\def\bar#1{\overline{#1}}
\def\bas{{\alpha}_s}
\def\bLambdabar{{{\overline \Lambda} \over 2 m_b}}
\def\bv{b_v}
\def\ccdot{\hbox{\kern-.1em$\cdot$\kern-.1em}}
\def\CD{{\cal D}}
\def\coupling{ {\bas(m_c) \over \pi}}
\def\cLambdabar{{{\overline \Lambda} \over 2m_c}}
\def\cvp{\overline{c}_{v'}}
\def\Dl{\overleftarrow{\CD}}
\def\Dlslash{{\overleftarrow{\CD}} \hskip-0.75 em / \hskip+0.40 em}
\def\Dlalpha{\overleftarrow{\CD}\null^\alpha}
\def\Dlmu{\overleftarrow{\CD}\null^\mu}
\def\Dslash{\CD\hskip-0.65 em / \hskip+0.30 em}
\def\GeV{\>\, \rm GeV}
\def\gfive{\gamma^5}
\def\gu{\gamma^\mu}
\def\hv{h_v^{{\scriptscriptstyle{(Q)}}}}
\def\hvbar{{\overline{h}_v^{{\scriptscriptstyle{(Q)}}}}}

\def\imb{\displaystyle{i \over m_b}}
\def\imc{\displaystyle{i \over m_c}}

\def\IW{\eta(v \ccdot v')}

\def\Lambdab{\Lambda_b(v,s)}
\def\Lambdac{\Lambda_c(v',s')}
\def\LB{\Lambda_b}
\def\LC{\Lambda_c}

\def\mQ{{m_{\scriptscriptstyle Q}}}

\def\OMIT#1{\null}
\def\Q{{\scriptscriptstyle Q}}


\def\ubar{\, {\overline u}(v',s')}
\def\u{u(v,s)}
\def\v{v^\mu}
\def\vp{v'^\mu}
\def\vsl{v \hskip-5pt /}
\def\vv{v \ccdot v'}


\def\fourthirds{{4 \over 3}}
\def\half{{1 \over 2}}
\def\third{{1 \over 3}}


\def\np#1#2#3{Nucl. Phys. {\bf #1} (#2) #3}
\def\pl#1#2#3{Phys. Lett. {\bf #1} (#2) #3}

\def\sjnp#1#2#3{Sov. J. Nucl. Phys. {\bf #1} (#2) #3}


\newdimen\pmboffset
\pmboffset 0.022em
\def\oldpmb#1{\setbox0=\hbox{#1}%
 \copy0\kern-\wd0
 \kern\pmboffset\raise 1.732\pmboffset\copy0\kern-\wd0
 \kern\pmboffset\box0}
\def\pmb#1{\mathchoice{\oldpmb{$\displaystyle#1$}}{\oldpmb{$\textstyle#1$}}
      {\oldpmb{$\scriptstyle#1$}}{\oldpmb{$\scriptscriptstyle#1$}}}


\nref\IsgurWise{N. Isgur and M.B. Wise, \pl{B232}{1989}{113};
  \pl{B237}{1990}{527}.}
\nref\Eichten{E. Eichten and B. Hill, \pl{B234}{1990}{511}.}
\nref\Georgi{H. Georgi, \pl{B240}{1990}{447}.}
\nref\Grinstein{B. Grinstein, \np{B339}{1990}{253}.}
\nref\Volopol{M.B. Voloshin and M.A. Shifman, \sjnp{45}{1987}{292}\semi
  H.D. Politzer and M.B. Wise, \pl{B206}{1988}{681};
  \pl{B208}{1988}{504}.}
\nref\FGGW{A. Falk, H. Georgi, B. Grinstein and M.B. Wise,
  \np{B343}{1990}{1}.}
\nref\Falk{A. Falk and B. Grinstein, \pl{B247}{1990}{406}.}
\nref\EichtenHill{E. Eichten and B. Hill, \pl{B243}{1990}{427}\semi
  M. Golden and B. Hill, \pl{B254}{1991}{225}\semi
  A.F. Falk and B. Grinstein, \pl{B249}{1990}{314}.}
\nref\FGL{A. Falk, B. Grinstein and M. Luke, \np{B357}{1991}{185}.}
\nref\Luke{M.E. Luke, \pl{B252}{1990}{447}.}
\nref\GGW{H. Georgi, B. Grinstein and M. B. Wise, \pl{252B}{1990}{456}.}
\nref\Boyd{C.G. Boyd and D. Brahm, Phys. Lett {\bf B257} (1991) 393.}
\nref\Lebed{R.F. Lebed and M. Suzuki, Phys. Rev. {\bf D44} (1991) 829.}
\nref\Wise{M.B. Wise, CALT-68-1721, Lectures presented at the Lake
  Louise Winter Institute.}
\nref\Baryonrefs{N. Isgur and M.B. Wise, \np{B348}{1991}{276}\semi
  H. Georgi, \np{B348}{1991}{293}\semi
  T. Mannel, W. Roberts and Z.Ryzak, \np{B355}{1991}{38};
  \pl{B255}{1991}{593}.}



%
\nfig\Graphs{One-loop Feynman diagrams in the intermediate and final HQET's
whose difference determines the $O(\bas(m_c))$ matching contributions to the
${C^{(3)}_j}^{(\prime)}$ and ${C^{(4)}_k}^{(\prime)}$ coefficients in
eqn.~\fulltoeff.  Solid boxes denote $P_0^{(\prime)}$ and $Q_1^{(\prime)}$
current operators while solid dots represent $O_1$ and $O_2$ Lagrangian
operator insertions.}



\def\LongTitle#1#2#3{\nopagenumbers\abstractfont\hsize=\hstitle\rightline{#1}%
\vskip 0.5in\centerline{\titlefont #2}\centerline{\titlefont #3}
\abstractfont\vskip 0.4in\pageno=0}
\LongTitle
  {\vbox{\hbox{SSCL--Preprint--111}\hbox{HUTP-92/A012}}}
  {Heavy Hadron Form Factor Relations} {for $m_c\ne\infty$ and $\bas(m_c)\ne0$}
\centerline{Peter Cho
  \footnote{$^\dagger$}{E-mail: Cho@huhepl.harvard.edu}}
\centerline{Lyman Laboratory of Physics}
\centerline{Harvard University}
\centerline{Cambridge, MA 02138}
\medskip\centerline{and}\medskip
\centerline{Benjam\'\i n Grinstein
  \footnote{$^\ddagger$}{E-mail: Grinstein@sscvx1.bitnet, @sscvx1.ssc.gov}
  \footnote{}{On leave of absence from Harvard University.}}
\centerline{Superconducting Super Collider Laboratory}
\centerline{2550 Beckleymeade Ave.}
\centerline{Dallas, TX 75237}

\vskip 0.6in


        First order power corrections to current matrix elements between
heavy meson or $\Lambda_\Q$ baryon states are shown to vanish at the zero
recoil point to all orders in QCD.  Five relations among the six form
factors that parametrize the semileptonic decay $\Lambda_b \to \Lambda_c e
\overline{\nu}$ are also demonstrated to exist to all orders in the strong
coupling at order $1/\mQ$ .  The $O(\bas(m_c)/m_c)$ form factor
relations are displayed.

\vskip 0.5 in
\centerline{Submitted to \it{Physics Letters B}}

\Date{April 1992} 
\newsec{Introduction}

        In the limit where the masses of the charm and bottom quarks are taken
to be infinitely greater than the QCD scale, matrix elements between hadron
states containing a single heavy quark are severely constrained \IsgurWise.
For example, all six form factors for the flavor changing currents which
mediate $B\to D$ and $B \to D^*$ transitions are given in terms of one
universal function. This so called ``Isgur-Wise'' function is also the form
factor of the $b$-number current between $B$ meson states.  It is
consequently normalized to unity at the maximum momentum transfers
$q^2_{\rm max}=0$ for $B\to B$ transitions and
$q^2_{\rm max}=(m_B-m_D)^2$ for $B\to D$ decays.

        A Heavy Quark Effective Theory (HQET) with manifest flavor and spin
symmetries that lead to these normalization constraints has recently been
developed \refs{\Eichten,\Georgi}.  Since the HQET is derived from
QCD \Grinstein,
its predictions are model independent.  Moreover, corrections to results
found in the infinite quark mass limit can be systematically investigated
in this effective theory.  Such corrections arise from QCD scaling
violations which depend logarithmically upon the charm and bottom
masses~\refs{\Volopol,\FGGW}.
In addition, terms suppressed by inverse powers of the
heavy quark masses enter at subleading order~\refs{\Falk,\EichtenHill,\FGL}.
We shall refer to these deviations from the infinite mass limit as ``scaling''
and ``power'' corrections respectively.

        First order power corrections to the predicted normalization of
flavor changing current matrix elements between $B$ and $D$ or
$D^*$ states have been shown to vanish at the zero recoil point \Luke.
This remarkable result is often called ``Luke's theorem'' and
holds as well for $\LB\to\LC$ transitions \GGW\
and for an entire class of heavy hadron processes \Boyd.
Luke's theorem was originally proved to zeroth order in the strong
interactions.  It consequently ruled out normalization corrections at
$O(1/m_c)$ but not $O(\as/m_c)$.  In this letter, we demonstrate that these
latter violations are also prohibited.  In fact, we show that
{\it there are no order 1/$m_c$ corrections to the zero recoil normalization
of the current matrix elements to all orders in $\as$.}

        We then focus our attention upon the semileptonic decay
$\LB\to\LC e \overline{\nu}$.  This process is of considerable interest
since an accurate value for the KM matrix element $|V_{cb}|$ may
be determined in the future from high precision measurements of its
endpoint spectrum.   The transition lends itself particularly well to HQET
analysis because it is tightly constrained by the heavy quark spin symmetry.
Like their mesonic counterparts, the six form factors that parametrize this
baryonic process are predicted at leading order in terms of a single
Isgur-Wise function. Five relations among these six form factors have been
found to remain after $O(1/m_c)$ power corrections are
included.  We extend this result to all orders in the strong coupling
and then display the relations to $O(\bas(m_c)/m_c)$.
Such form factor relations provide a valuable means for assessing the
uncertainty in future measurements of the mixing angle $|V_{cb}|$
from semileptonic $\Lambda_b$ decay.

        Finally, we estimate and compare the numerical sizes of the scaling and
power correction expansion parameters that appear in the HQET.

\newsec{Nonrenormalization at the zero recoil point}

        Finite quark mass corrections enter into the HQET in two
ways.  Firstly, $O(1/m_c)$ and $O(1/m_b)$ terms appear in the Lagrangian
which break the theory's flavor and spin symmetries:
\eqn\Lagrangian{
\CL_v = \sum_{\Q=c,b} \Bigl\{ \hvbar (iv \ccdot \CD) \hv + a_1 O_1 + a_2 O_2
  \Bigr\}. }
The $O_i$ operators are built up out of two heavy quark fields and
symmetric or antisymmetric combinations of two covariant derivatives
$\CD_\mu = \partial_\mu - i g A^a_\mu T^a$:
\foot{A third operator $O_3=-(1/2\mQ) \hvbar (i v \ccdot \CD)^2 \hv$ could
be included with those in \Oops.  However, since it can be eliminated via
a nonlinear field redefinition, this operator has no effect and can be
neglected without loss \FGL.}
%
%
\eqn\Oops{\eqalign{
O_1 &= {1 \over 2\mQ} \hvbar (i\CD)^2 \hv \cr
O_2 &= {g \over 4\mQ} \hvbar \sigma^{\mu\nu} G^a_{\mu\nu} T^a \hv .\cr}}
We have absorbed various numerical factors into these operators'
definitions so that their tree level coefficients equal unity:
\eqn\acoeffs{a_1 = a_2 = 1+O(\bas).}
The Ademollo-Gatto theorem indicates that corrections to the normalization of
form factors from the $O_i$ terms in \Lagrangian\ arise only at second order
in $1/m_c$ and $1/m_b$ \refs{\Boyd,\Lebed}.
The QCD corrections to the $a_i$ coefficients in \acoeffs\ do not upset this
result.

        There are also power corrections to the effective currents in the HQET
which correspond to the vector and axial currents in the underlying full
theory.  In general, the two sets of currents are related as
\eqn\fulltoeff{\eqalign{
V^\mu &= \overline{c} \gamma^\mu b \to \sum C^{(3)}_j P^\mu_j
  + \sum C^{(4)}_k Q^\mu_k + \cdots\cr
A^\mu &= \overline{c} \gamma^\mu\gamma^5 b \to \sum {C^{(3)}_j}' P'^\mu_j +
  \sum {C^{(4)}_k}' Q'^\mu_k +\cdots . \cr}}
Here $P_j^{(\prime)\mu}$ and $Q_k^{(\prime)\mu}$ denote dimension three and
four
operators with appropriate quantum numbers while the ellipses represent higher
order terms.  A convenient basis for these operators is listed below:
\eqna\PQops
\smallskip\noindent${\underline{{\rm Dimension} \; 3:}}$
$$ \vbox{\settabs \+ \quad\qquad\qquad & \qquad\qquad &
  $P^\mu_0$ & $=$ & $\cvp \gu \bv$
  \qquad \qquad \qquad & $P'^\mu_0$ & $=$ & $\cvp \gu \gfive \bv$ \quad \cr
\+ \quad \qquad \qquad & \qquad \qquad &
  \hfill $P^\mu_0$ & $=$ & $ \cvp \gu \bv$ \hfill &
  \hfill $P'^\mu_0$ & $=$ & $ \cvp \gu\gfive \bv$ \hfill \quad \cr
\+ \quad \qquad \qquad & \qquad \qquad &
  \hfill $P^\mu_1$ & $=$ & $ \cvp \v \bv$ \hfill &
  \hfill $P'^\mu_1$ & $=$ & $ \cvp \v \gfive \bv$ \hfill
  $\>$ \quad \qquad (2.5a)  \cr
\+ \quad \qquad \qquad & \qquad \qquad &
  \hfill $P^\mu_2$ & $=$ & $ \cvp \vp \bv$ \hfill &
  \hfill $P'^\mu_2$ & $=$ & $ \cvp \vp \gfive \bv$ \hfill \quad \cr} $$
\noindent${\underline{{\rm Dimension} \; 4:}}$
$$ \vbox{\settabs \+ \qquad\qquad \qquad &
$Q_{13}$ & $=$ &  $\> -\imc \cvp v\ccdot \Dl \vp \bv \>$
\qquad\qquad & $Q'_{13}$ & $=$ & $\> -\imc \cvp \v\ccdot \Dl \vp \gfive \bv \>$
\cr
\+ \qquad\qquad \qquad & \hfill $Q^\mu_1$ &
$=$ & $ -\imc \cvp \Dlslash \gu \bv$
 \hfill &
  \hfill $Q'^\mu_1$ & $=$ & $ -\imc \cvp \Dlslash \gu\gfive \bv$ \hfill \cr
\+ \qquad\qquad \qquad & \hfill $Q^\mu_2$ &
$=$ & $ \imb \cvp \gu \Dslash \bv$ \hfill &
  \hfill $Q'^\mu_2$ & $=$ & $ \imb \cvp \gu\gfive \Dslash \bv$ \hfill \cr
\+ \qquad\qquad \qquad & \hfill $Q^\mu_3$ &
$=$ & $ -\imc \cvp v\ccdot\Dl \gu \bv$ \hfill &
  \hfill $Q'^\mu_3$ & $=$ & $ -\imc \cvp v\ccdot\Dl \gu\gfive \bv$ \hfill \cr
\+ \qquad\qquad \qquad & \hfill $Q^\mu_4$ &
$=$ & $ \imb \cvp \gu v'\ccdot \CD \bv$ \hfill &
  \hfill $Q'^\mu_4$ & $=$ & $ \imb \cvp \gu\gfive v'\ccdot \CD \bv$ \hfill \cr
\+ \qquad\qquad \qquad & \hfill $Q^\mu_5$ & $=$ & $ -\imc \cvp \Dlslash \v \bv$
\hfill &
  \hfill $Q'^\mu_5$ & $=$ & $ -\imc \cvp \Dlslash \v \gfive \bv$ \hfill \cr
\+ \qquad\qquad \qquad & \hfill $Q^\mu_6$ &
$=$ & $ -\imc \cvp \Dlslash \vp \bv$
 \hfill &
  \hfill $Q'^\mu_6$ & $=$ & $ -\imc \cvp \Dlslash \vp \gfive \bv$ \hfill \cr
\+ \qquad\qquad \qquad & \hfill $Q^\mu_7$ &
$=$ & $ \imb \cvp \v  \Dslash \bv$ \hfill &
  \hfill $Q'^\mu_7$ & $=$ & $ \imb \cvp \v\gfive \Dslash \bv$ \hfill
  $\>$ ~~~ (2.5b)\cr
\+ \qquad\qquad \qquad & \hfill $Q^\mu_8$ &
$=$ & $ \imb \cvp \vp \Dslash \bv$ \hfill &
  \hfill $Q'^\mu_8$ & $=$ & $ \imb \cvp \vp \gfive \Dslash \bv$ \hfill \cr
\+ \qquad\qquad \qquad & \hfill $Q^\mu_9$ &
$=$ & $ -\imc \cvp \Dlmu \bv$ \hfill
 &
  \hfill $Q'^\mu_9$ & $=$ & $ -\imc \cvp \Dlmu \gfive \bv$ \hfill \cr
\+ \qquad\qquad \qquad & \hfill $Q^\mu_{10}$ &
$=$ & $ \imb \cvp \CD^\mu \bv$ \hfill &
  \hfill $Q'^\mu_{10}$ & $=$ & $ \imb \cvp \gfive \CD^\mu \bv$ \hfill \cr
%
%
\+ \qquad\qquad \qquad & \hfill $Q^\mu_{11}$ &
$=$ & $ -\imc \cvp v\ccdot \Dl \v
 \bv$ \hfill &
  \hfill $Q'^\mu_{11}$ & $=$ & $ -\imc
 \cvp v\ccdot \Dl \v \gfive \bv$ \hfill \cr
\+ \qquad\qquad \qquad & \hfill $Q^\mu_{12}$ &
$=$ & $ -\imc \cvp v\ccdot \Dl \vp \bv$ \hfill &
  \hfill $Q'^\mu_{12}$ & $=$ &
$ -\imc \cvp v\ccdot \Dl \vp\gfive \bv$ \hfill \cr
\+ \qquad\qquad \qquad & \hfill $Q^\mu_{13}$ &
$=$ & $ \imb \cvp \v v'\ccdot \CD \bv$ \hfill &
  \hfill $Q'^\mu_{13}$ & $=$ &
$ \imb \cvp \v\gfive v'\ccdot \CD \bv$ \hfill \cr
\+ \qquad\qquad \qquad & \hfill $Q^\mu_{14}$ & $=$ & $ \imb
 \cvp \vp v'\ccdot \CD \bv$ \hfill &
  \hfill $Q'^\mu_{14}$ & $=$ & $ \imb \cvp \vp\gfive v'\ccdot \CD \bv$.\hfill
  \cr}$$
The operators' coefficients are determined by matching Green's functions
with single current insertions in the full and effective theories.
They are dimensionless functions of the strong coupling $\as$, the
renormalization point $\mu$, and the quark masses $m_c$ and $m_b$.
Their values can be calculated perturbatively provided
$\mu$ is large enough so that $\as(\mu)$ is small.

        All of the effective current operator coefficients in \fulltoeff\
gain zero contribution from tree level matching except
$$ \eqalign{C^{(3)}_0 &= {C^{(3)}_0}'=1 \cr
C^{(4)}_1 &= {C^{(4)}_1}' = C^{(4)}_2 = {C^{(4)}_2}'=1/2.\cr} $$
In the original proof of Luke's theorem, only the
operators corresponding to these nonvanishing coefficients were
considered.  To extend the theorem's validity to arbitrary order in $\as$, one
must examine the effects from all the others listed in \PQops{}.
Therefore, consider a representative HQET matrix element of a prototype
dimension four operator between heavy $B$ and $D$ states that both move with
four-velocity $v$:
\eqn\firstidentity{
\vev{\widetilde D(v)|\overline{c}_{v} i\Dl{}^\alpha \Gamma b_v |
  \widetilde B(v)} = \lambda v^\alpha \Tr \overline{M}(v)\Gamma  M(v).}
%
The tildes appearing on the LHS of this equation indicate that the states are
evaluated in the effective theory to zeroth order in $1/\mQ$.  On the RHS, the
meson matrices
$$ \eqalign{M(v) &= -{1 +\vsl \over 2} \gfive \cr
\overline{M}(v) &= \gfive {1+\vsl \over 2} \cr} $$
%
are contracted together in accordance with the HQET flavor and spin symmetries.
After dotting both sides of \firstidentity\ with $v_\alpha$ and applying
the equation of motion $v\ccdot \CD c_v=0$, one finds that the constant
$\lambda$ vanishes identically.  Since matrix elements between $B(v)$ and
$D(v')$ states of all the dimension four operators in \PQops{b}\
can be derived from equations like \firstidentity, they too must vanish
when $v=v'$.  An analogous argument holds for $B\to D^*$ transitions.

        Could the zeros in heavy meson matrix elements of the
$Q_k^{(\prime)\mu}$ operators be cancelled by poles in their
${C_k^{(4)}}^{(\prime)}$ coefficients?  We do not believe so.  Consider the
analytic structure of meson form factors regarded as complex functions of
the momentum transfer $q^2$.  By examining Feynman diagrams in the underlying
full QCD theory,  one sees that the physical cut which starts at the maximum
momentum transfer $q^2_{\rm max}=(m_B - m_D)^2$ originates from infrared
singularities in these graphs.  This infrared behavior must be reproduced by
the dynamics of the effective theory and should not appear in the coefficient
functions which contain only short distance information.

        Therefore, since matrix elements of the dimension four operators
vanish while their coefficients remain regular at $\vv=1$, there can
be no first order power corrections to the zero recoil current normalizations
to all orders in QCD.

\newsec{Form factor relations for $\pmb{\LB\to\LC}$ transitions}

        The nonrenormalization theorem discussed in the previous section
for mesons applies to $\Lambda_\Q$ baryons as well.  Vector and axial current
matrix elements between $\LB$ and $\LC$ baryon states appear in the HQET as
\eqn\ffsdefn{\eqalign{
\vev{\Lambdac|V^\mu|\Lambdab} &= \ubar[F_1(\vv) \gamma^\mu + F_2(\vv) v^\mu
  + F_3(\vv) v'^\mu ]\u \cr
\vev{\Lambdac|A^\mu|\Lambdab} &= \ubar [G_1(\vv) \gamma^\mu + G_2(\vv) v^\mu
  + G_3(\vv) v'^\mu ]\gfive \u. \cr}}
A few points about these expressions should be noted.  Firstly, the Dirac
spinors for the baryons' heavy quark constituents satisfy $\u =\vsl\u$.
Therefore when $v = v'$, the current matrix elements reduce to \Wise
\eqna\reducedelems
$$ \eqalignno{
\vev{\Lambda_c(v,s')|V^\mu|\Lambdab} &= [F_1(1)+F_2(1)+F_3(1)]
  \overline{u}(v,s')v^\mu \u &\reducedelems a \cr
\vev{\Lambda_c(v,s')|A^\mu|\Lambdab} &= G_1(1) \overline{u}(v,s')
  \gamma^\mu \gfive \u. & \reducedelems b\cr} $$
Secondly, the spin of a $\Lambda_\Q$ baryon comes entirely from
its heavy quark in the infinite mass limit; the light spectator degrees
of freedom carry zero angular momentum.  The form
factors $F_i$ and $G_i$ are consequently all determined from
one universal function which is normalized at zero recoil \Baryonrefs.
To avoid any confusion with the Isgur-Wise function $\xi(\vv)$ for heavy
mesons, we will denote this universal function associated with $\Lambda_\Q$
baryons as $\IW$.  Finally, an additional dimensionful constant
$\overline{\Lambda} \approx m_{\Lambda_c}-m_c \approx m_{\Lambda_b}-m_b$
must be introduced to specify the form factors when $\mQ \ne\infty$.
The parameter $\overline{\Lambda}$ may be interpreted as the baryon
state's energy above the vacuum in the HQET.

        Order $1/m_c$ power corrections to the effective vector and axial
currents arising from either local dimension four $Q_k^{(\prime)\mu}$
operators in \PQops{b}\ or time ordered products of dimension five $O_i$
operators in \Lagrangian\ and dimension three $P_j^{(\prime)\mu}$ operators
in \PQops{a}\ were considered in ref.~\GGW.
The time ordered products were shown to generally not contribute, and five
relations among the six form factors in \ffsdefn\ were found.
We now demonstrate that five relations remain even when current corrections of
order $1/m_c$, $1/m_b$ and all orders in $\as$ are retained.
We start with the identity
$$ \vev{\Lambdac \vert i \CD^\alpha (\cvp \Gamma \bv) \vert \Lambdab} =
  \bar{\Lambda} \IW (v^\alpha - v'^\alpha) \ubar \Gamma \u $$
which follows from the relation between momentum and derivative operators
in the effective theory \Georgi:
$$ [ {\bf P^\alpha}, \hv(x) ] = -(m v^\alpha + i \CD^\alpha) \hv(x).$$
With the aid of this identity, the general matrix elements
\eqn\secondidentity{\eqalign{
\vev{\Lambdac \vert \cvp i \Dlalpha \Gamma \bv \vert \Lambdab} &=
  \bar{\Lambda} \IW {v^\alpha - v\ccdot v' v'^\alpha \over v \ccdot v'+1}
  \ubar \Gamma \u \cr
\vev{\Lambdac \vert \cvp \Gamma i \CD^\alpha \bv \vert \Lambdab} &=
  \bar{\Lambda} \IW {\vv v^\alpha - v'^\alpha \over v\ccdot v'+1 }
  \ubar \Gamma \u \cr}}
are readily evaluated.  Notice that like
the meson element in \firstidentity, these expressions vanish for
$v = v'$.

        Matrix elements of all the basis operators in \PQops{b}\
are fixed by those in \secondidentity.  Since any
dimension four contribution to the effective currents can be decomposed over
this complete operator set, we see that Luke's theorem holds to all powers
in the strong coupling.  Furthermore, as no new parameters
need be introduced into the current form factors, no relations among them
are lost.
%
%
%
%
%
%
Such relations can be determined to the order at which the
effective current coefficients in \fulltoeff\ are known.  We compute these
coefficients assuming $m_b \gg m_c$, and we first work in an intermediate
HQET with a heavy $b$ quark but full theory $c$ field.  For simplicity, we
neglect the QCD running between the bottom and charm scales which has
previously been discussed in refs.~\refs{\Volopol,\FGGW,\Falk}.
We instead concentrate upon the $O(\bas(m_c)/m_c)$ matching contributions to
the current coefficients that arise
at the charm scale boundary between the intermediate and final
effective theories in which both the $c$ and $b$ are treated as heavy.

        We match 1PI two-point Green's functions with a single vector or
axial current insertion in the intermediate and final HQET's.  The one-loop
diagrams that enter into this matching computation are illustrated in
\Graphs.  The graphs contain $O(1/m_c)$ operator insertions from the
Lagrangian in \Lagrangian\ and currents in \fulltoeff.
We adopt the mass independent renormalization scheme of dimensional
regularization plus modified minimal subtraction to accommodate the
ultraviolet infinities in these diagrams.  Infrared
divergences which appear after Taylor expanding loop integrals in powers of
external residual momenta can be explicitly eliminated by judiciously
arranging integrand terms in the two theories into infrared safe
combinations.  After including the tree level $O(1/m_c)$ and $O(1/m_b)$
contributions and taking the difference between the two-point functions in the
intermediate and final HQET's, we find the following $c$-scale matching
contributions to the effective currents:
\eqna\matchcurrents
$$\eqalign{
V^\mu &= P^\mu_0+\half Q^\mu_1 +\half Q^\mu_2 \cr
&+ \third \coupling \biggl \{ 2(\vv+1)r \bigl[P^\mu_0+\half Q^\mu_1 \bigr]
  -2 {r-1 \over \vv-1} Q^\mu_3 \cr
&\qquad\qquad\qquad -4r \bigl[P^\mu_1+\half Q^\mu_5 \bigr]
  -{4(1-\vv r) \over \vv^2-1} Q^\mu_{11} \biggr \} \cr}\eqno\matchcurrents a $$
\bigskip
$$ \eqalign{
A^\mu &= P'^\mu_0+\half Q'^\mu_1 + \half Q'^\mu_2 \cr
&+ \third \coupling \biggl \{ 2(\vv-1)r \bigl[P'^\mu_0+\half Q'^\mu_1 \bigr]
  +2 {r+1 \over \vv+1} Q'^\mu_3 \cr
&\qquad\qquad\qquad -4r \bigl[P'^\mu_1+\half Q'^\mu_5 \bigr]
  -{4(1-\vv r) \over \vv^2-1} Q'^\mu_{11} \biggr \} \cr}\eqno\matchcurrents b$$
where
$$ r = {\log(\vv +\sqrt{\vv ^2-1}) \over \sqrt{\vv ^2-1}}~.$$
The $O(\bas(m_c))$ coefficients of the dimension three terms in these formulas
are consistent with results from previous matching computations \FGGW.

        Five independent relations among the vector and axial
form factors are readily derived from the currents in
\matchcurrents{}.  We choose to express these relations as ratios relative to
the first axial form factor:
\def\myif#1#2{{\ifx\answ\bigans{#1}\else{#2}\fi}}
\eqna\ratios
\myif{$$ \eqalignno{
{F_1 \over G_1} &= 1+\Bigl[ \cLambdabar + \bLambdabar \Bigr]
  {2 \over (\vv+1)}  + \fourthirds\coupling r
  + \fourthirds\coupling\cLambdabar
  {2(1+r-\vv r) \over (\vv+1) } \quad &\ratios a \cr
{F_2 \over G_1} &= {G_2 \over G_1} =
  - \cLambdabar {2 \over (\vv+1) } - \fourthirds \coupling r
  - \fourthirds\coupling \cLambdabar {2(1+r-\vv r) \over (\vv+1) }
  \quad &\ratios b \cr
{F_3 \over G_1} &= -{G_3 \over G_1} =
  - \bLambdabar {2 \over (\vv+1) }. \quad &\ratios c \cr} $$}%
{$$ \eqalignno{
{F_1 \over G_1} &= 1+\Bigl[ \cLambdabar + \bLambdabar \Bigr]
  {2 \over (\vv+1)}  \cr
   &\qquad\qquad + \fourthirds\coupling r
  + \fourthirds\coupling\cLambdabar
  {2(1+r-\vv r) \over (\vv+1) } \quad &\ratios a \cr
{F_2 \over G_1} &= {G_2 \over G_1} =
  - \cLambdabar {2 \over (\vv+1) } \cr
   &\qquad\qquad - \fourthirds \coupling r
  - \fourthirds\coupling \cLambdabar {2(1+r-\vv r) \over (\vv+1) }
  \quad &\ratios b \cr
{F_3 \over G_1} &= -{G_3 \over G_1} =
  - \bLambdabar {2 \over (\vv+1) }. \quad &\ratios c \cr} $$}%
As a check, we have verified that these form factor relations are
renormalization scheme independent as must be the case for physical
observables.

        The ratios in \ratios{}\ imply $F_1 + F_2 + F_3 = G_1$
for all values of $\vv$.  Our enhanced version of Luke's theorem applied to
eqn.~\reducedelems{}\ guarantees that no dimension four terms disrupt the
normalization of these form factor combinations at zero recoil.
Possible normalization violations from dimension three terms are also
prohibited when $v=v'$ as can be readily verified in $v\ccdot A^a=0$
gauge.  Therefore to leading order, only calculable QCD scaling
corrections move the values
of these form factor combinations away from unity at the zero recoil
point:
\eqn\formcombos{
 F_1(1)+F_2(1)+F_3(1) = G_1(1) = \Bigl[ {\bas(m_b) \over
  \bas(m_c)} \Bigr]^{- 6/25} . }

        To conclude, we estimate the numerical sizes of the expansion
parameters that enter into HQET computations when the bottom and charm
quarks are sequentially treated as heavy and the running between them is
neglected.  Such calculations are organized as perturbative expansions in
$\bar{\Lambda}/2m_c$, $\as(m_c)/\pi$, $\bar{\Lambda}/2m_b$ and $\as(m_b)/\pi$.
Assuming the reasonable values $m_b=4.5 \GeV$, $m_c=1.5 \GeV$,
$\bar{\Lambda}=0.5 \GeV$ and $\Lambda_{\rm QCD}^{(3)}=0.2 \GeV$ and using the
leading log approximation for the strong interaction fine structure constant,
we find that the charm scale parameters $\bar{\Lambda}/2m_c=0.17$ and
$\as(m_c)/\pi=0.11$ are of comparable magnitude.  Their squares
$(\bar{\Lambda}/2m_c)^2=0.03$, $(\as(m_c)/\pi) \bar{\Lambda}/2m_c =0.02$ and
$(\as(m_c)/\pi)^2=0.01$ are not much
smaller than the bottom scale expansion parameters $\as(m_b)/\pi=0.07$ and
$\bar{\Lambda}/2m_b=0.05$.  Further corrections lie below the 1 \% level.  The
uncertainty in the relations \ratios{}\ and \formcombos\ is therefore
dominated by second order $(\bar{\Lambda}/2m_c)^2$ power corrections.  Such
terms are comparable in size to the order $(\as(m_c)/\pi)
\bar{\Lambda}/2m_c$ contributions that we have considered here.

\bigskip
\noindent{\bf Acknowledgements}
\bigskip

        It is a pleasure to acknowledge helpful discussions with Howard Georgi
and Mark Wise.  BG would like to thank the Alfred P. Sloan
Foundation for partial support. This work  was supported in part by the
National Science Foundation under grant  PHY--87--14654, by the
Texas National Research Laboratory Commission
under grant \#RGFY9106, and by the Department of Energy under contract
DE--AC35--89ER40486.

\listrefs
\vfill\eject\immediate\closeout\ffile{\parindent40pt
\baselineskip14pt\centerline{{\bf Figure Caption}}\nobreak\medskip
\escapechar=` \input figs.tmp\vfill\eject}
\bye